\documentclass{ws-m3as}

\usepackage{amssymb}
\usepackage{amsmath}
\usepackage{enumerate}
\usepackage{amstext}
\usepackage{graphicx}
\usepackage{hhline}
\usepackage{psfrag}
\usepackage{float}
 \usepackage{rotating}
\usepackage{subfig}
\usepackage{epstopdf}
 \usepackage{graphics}
\usepackage{tikz}

 \newcommand{\vectornorm}[1]{\left|\left|#1\right|\right|}

\title{Numerical simulation of moving rigid bodies in  rarefied gases}

\author{Samir Shrestha}
\address{
Department of Mathematics,
University of Kaiserslautern, \\ P.O.Box 3049, 67653
Kaiserslautern, Germany.\\
shrestha@mathematik.uni-kl.de
}

\author{Sudarshan Tiwari}
\address{
Department of Mathematics,
University of Kaiserslautern, \\ P.O.Box 3049, 67653
Kaiserslautern, Germany.\\
tiwari@mathematik.uni-kl.de
}
\author{Axel  Klar}
\address{
Department of Mathematics,
University of Kaiserslautern, \\ P.O.Box 3049, 67653
Kaiserslautern, Germany.\\ Fraunhofer ITWM Kaiserslautern, 67663 Kaiserslautern, Germany.\\
klar@mathematik.uni-kl.de
}
 \author{Steffen Hardt}
 \address{Center of Smart Interfaces, TU Darmstadt\\
Alarich-Weiss-Str. 10,, 64287, TU Darmstadt Germany.\\
hardt@csi.tu-darmstadt.de
}
  
\begin{document}
\maketitle

\begin{abstract}
In this paper we present a numerical scheme to simulate a moving rigid body with arbitrary shape suspended in a rarefied gas. 
The rarefied gas is simulated by solving the Boltzmann equation using a DSMC particle method. 
The motion of the rigid body is governed by the Newton-Euler equations, 
where the force and the torque on the rigid body is computed from the momentum transfer of the gas molecules colliding with the  body. 
On the other hand, the motion of the rigid body influences  the gas flow in its surroundings. 
We validate the numerical results by testing the Einstein relation for Brownian motion of the suspended particle. 
The translational as well as the rotational degrees of freedom are taken into account. 
It is shown that the numerically computed translational and rotational diffusion coefficients converge to the  
theoretical values. 
\end{abstract}

Key Words \texttt{rigid body motion, Boltzmann equation, DSMC, moving body in rarefied gas, Brownian diffusion}
MSC 2010 65C05; 65C30; 74F10



 \section{Introduction}
 Nanoparticles play a vital role in many industrial processes and natural phenomena, including areas such as chemical 
 engineering, biomedical technology, material science, physics, chemistry, and biology. 
Nanoparticles are often suspended in fluids during production, handling, and processing, but also after  unintentional or undesired release to the environment. In many cases the suspending fluid is a gas, as 
 in the case of large scale commercial production of nanoparticles, air pollution or clean room technology. 
 
 These applications have increased the interest in micro- and nanofluidics and have  triggered the development of simulation methods and numerical schemes.
 For example,  Direct Simulation Monte Carlo (DSMC) methods have been used in the modeling of small-scale devices with characteristic dimensions of microns down to nanometers \cite{KBA}. Furthermore, DSMC is an ideal particle based scheme for the study of hydrodynamic fluctuations \cite{balakrishnan}.
Moving boundary problems for kinetic equations have been recently investigated extensively  in connection with MEMS, see \cite{KBA} for an overview. To simulate such problems, DSMC as well as deterministic methods have been applied. In recent years several works  have  been reported for moving boundary 
problems in  a rarefied gas, see \cite{GRT}, \cite{RF}, \cite{TA}, \cite{DM}. 
To simulate moving rigid bodies  in a rarefied gas, not only the influence of the moving boundary on the gas has to be included 
in the simulation, but also the forces exerted by the gas  accelerating the rigid body. See, for example \cite{TA,TKHD} for  
one-dimensional situations with such a two-way coupling. 
We remark that, on the one hand, using  DSMC based approaches for the above time-dependent problems with slow fluid flows
requires some control over the  large fluctuations inherent in these methods. On the other hand,  the deterministic approaches are complicated to extend and computationally costly for  higher dimensions.  Finally, we note that  DSMC 
methods are especially suited to couple moving rigid objects due to the Lagrangian nature of the gas molecules. 

In this paper we present a simulation scheme for moving rigid bodies of arbitrary shape suspended in  a rarefied gas suited for three dimensions. As an implementation of the scheme we consider 
a rigid spherical body contained inside a cube of micron size.  The rarefied gas flow is simulated  
by solving the Boltzmann 
equation in a time dependent domain of computation using a DSMC particle method. The rigid body motion is given  by the Newton-Euler equations, 
where the forces on the rigid body are calculated from the momentum transfer due to gas molecules impinging on the surface of the rigid body. The resulting motion of the  rigid body affects in turn again  the gas flow in the surroundings. 
This means that a two-way coupling has to be modeled.  
To validate our numerical scheme, we investigate 3D Brownian motion of a spherical rigid body suspended in a gas 
and compare the numerical results with the Einstein relation. Both the translational 
and the rotational degrees of freedom of the particle are taken into account. Furthermore, we perform a convergence study comparing the 
numerically computed  translational and rotational diffusion coefficients to existing theoretical values  for an 
increasing number of simulated gas molecules. 
   
The paper is organized as follows. In section 2, we briefly present the Boltzmann equation and a numerical 
method for solving it. 
In section 3, we derive the  force and the torque exerted on the rigid body due to the surrounding gas medium and 
also present a  scheme based on the interaction of gas molecules and the rigid body and its implementation to estimate the 
force and the torque on the rigid body in a DSMC framework. 
In section 4, we describe the translational and rotational Brownian motions of a rigid spherical particle suspended in a 
gas  based on  Langevin's equations and also present the derivation of the corresponding diffusion coefficients.  
The numerical results are presented in section 5, and some concluding remarks are given in section 6.

\section{The Boltzmann equation}
  \subsection{The  equation}
The Boltzmann equation describes the time evolution of a distribution function $f(t,{\bf x},{\bf v})$ for particles with 
velocity ${\bf v} \in {\mathbb R}^d, d = 1, 2, 3$  and position ${\bf x}\in {\mathbb R}^3$ at time $t>0$. 
It is given in nondimensional form as \cite{}
\begin{equation}\label{Eq:Boltzmann}
\frac{\partial f}{\partial t} + {\bf v}\cdot{\boldsymbol\nabla_x} f = \frac{1}{\epsilon} J(f, f),
\end{equation}
with the initial condition 
\begin{equation}\label{Eq:BoltzmannInitialCondition}
f(t=0,{\bf x},{\bf v}) = f_0({\bf x},{\bf v}), 
\end{equation}
where $\epsilon$ is the Knudsen number, the ratio of mean free path and characteristic length, $J(f,f)$ is the 
collision operator which is given for hard-sphere molecules by
\begin{equation}\label{Eq:BoltzmannCollisionOperator}
J(f, f) = \int_{\mathbb{R}^3}\int_{S^2} \beta(\vectornorm{\bf v-w},{\bf n})[f({\bf v}')f({\bf w}')-f({\bf v})f({\bf w})] d\omega({\bf n})d{\bf w},
\end{equation}
where $S^2$ is the unit sphere in $\mathbb{R}^3$, ${\bf n}\in S^2$ is the unit vector in the impact direction, $\beta$ is the 
collision cross section, $f({\bf v}')=f(t,{\bf x},{\bf v}')$, and analogously for $f({\bf v})$ etc. The pair $({\bf v},{\bf w})$ and  $({\bf v}',{\bf w}')$ 
are the pre- and post- collisional velocities of two colliding gas molecules, given by 
\begin{equation}\label{Eq:BoltzmannPrePostVelocity}
{\bf v}' = {\bf v-n[n\cdot(v-w)]}, {\bf w}' = {\bf w + n[n\cdot(v-w)]}.
\end{equation} 
For more details we refer to \cite{cercignani1}. 
We note that in this paper the characteristic length is the diameter of the rigid body.
For the investigations in the present paper one has to solve (\ref{Eq:Boltzmann}) in  a time dependent domain given by the moving nanoparticle and predefined outer boundaries.
We consider  diffuse reflection boundary conditions at the outer boundaries of the domain of computation
and 
 at the surface of the rigid body. 
It is worth noting that the diffuse reflection boundary condition at the  
rigid body has to be evaluated in the co-moving frame of reference. 
 
\subsection{Numerical method for the Boltzmann equation}

We solve the Boltzmann equation using  
a variant of the DSMC method \cite{bird}, developed in 
\cite{NS95}, \cite{BI89}. The method is based on 
the time splitting of the Boltzmann equation. Introducing 
fractional steps one first solves the free transport equation (the 
collisionless Boltzmann equation) for one time step. During the 
free flow,  boundary conditions are taken into account. 
In this paper we consider a closed cube containing a rigid spherical body. 
As already mentioned, diffuse reflection boundary conditions apply at all boundaries.  
In a second step (the collision step), the spatially 
homogenous Boltzmann equation 
without the transport term is solved. 
To solve the homogeneous Boltzmann equation, the key point is to find 
an efficient particle approximation of the product distribution functions 
in the Boltzmann collision operator given only an approximation of the 
distribution function itself. To simulate this equation by a 
particle method an explicit Euler step is performed. 
To guarantee positivity of the distribution 
function during the collision step, a restriction of the time step 
proportional to the Knudsen number is needed. That means that the 
method becomes exceedingly expensive for small Knudsen numbers. 
In such regimes, a special algorithm adapted to the small Knudsen number limit 
has to be used, see, e.g., \cite{TKH}. 
Since here we aim at the large Knudsen number regime, we will not 
go into details concerning this issue.

The cube is discretized using a uniform grid size along all axes, resulting in a cubic grid. 
The initial phase space distribution of the gas is a Maxwellian distribution with the initial temperature, density and mean velocity as its parameters. 
The temperature is kept at the initial 
temperature throughout the simulations. The initial mean velocity and the initial velocity of the rigid 
body is zero. 
 
The computational grid is divided into three sets. Gas cells 
 completely filled by gas molecules, rigid body cells completely covered by the rigid body and boundary cells which are partially filled by gas molecules 
 and partially  by the rigid body. 
We note that due to the motion of the body we  have to update the  volume of the  cells occupied by the gas  at every time step.  
This update can be done efficiently  by marking the boundary cells and its neighboring cells near the surface of the rigid body.  Only the boundary cells and 
its neighbor cells are the candidate of the boundary cells in the next time step. 
 Some computational efforts are necessary to update the volume of boundary cells occupied by the gas. This is obtained by a 
Monte Carlo method using particles. Let $n_0$ be the initial number of gas molecules per cell. We distribute randomly $n_0$ molecules in 
boundary cells and we set a counter how many molecules are lying outside the rigid body. We repeat this process for $500$ times and then take the 
average of the counter. 
The gas volume fraction and the direction of the surface normal of the rigid body in the boundary cells are sufficient for the DSMC simulation 
near the boundary.

\section{Force and torque on the rigid body}

Generally, a rigid body suspended in a gas (representing, for example, a  nanoparticle) moves under the influence of surface and volume forces. The surface forces are due to collisions of gas molecules with the body. In addition, different types of volume forces could be present, such as electrostatic and gravitational forces. In this work, we only consider the effect of the surface forces onto the rigid body. It should be noted,  that via the solution of the Newton-Euler equations, inertial forces are taken care of automatically. 

\subsection{Force, torque and equation of motion}

To compute the surface force exerted on the rigid body by the surrounding gas we proceed as follows. 
We note that the following process can be applied to an arbitrarily  shaped rigid body with obvious modifications. In 
this paper we consider a rigid spherical  particle. 
Let $S(t)=\{{\bf y}(t):\vectornorm{{\bf y}(t)-{\bf X}(t)}\le R_\textrm{\tiny P}\}$ be  the rigid 
spherical particle with boundary $\partial S(t) = \{{\bf y}(t):\vectornorm{{\bf y}(t)-{\bf X}(t)}=R_\textrm{\tiny P}\}$ and 
center of mass ${\bf X}(t)$ at any moment in time $t$. The force $\pmb{\mathcal{{F}}}$ and the torque $\pmb{\mathcal{T}}$ 
exerted on the rigid sphere from the surrounding fluid is given by 
\begin{eqnarray}
\pmb{\mathcal{{F}}} &=& -\int_{\partial S} \sigma\cdot{\bf n}_{s} dA \label{Eq:HydrodynamicForceOnSphere} \\
\pmb{\mathcal{T}}&=& -\int_{\partial S} ({\bf y}-{\bf X})\times(\sigma\cdot{\bf n}_{s}) dA \label{Eq:HydrodynamicTorqueOnSphere}, 
\end{eqnarray}
where $\sigma$ is the total stress tensor in the fluid and ${\bf n}_s$ is the outward normal to the 
boundary $\partial S$ of the body. Using equations (\ref{Eq:HydrodynamicForceOnSphere}) 
and (\ref{Eq:HydrodynamicTorqueOnSphere}), the translational and rotational motion of the rigid 
body is described by the Newton-Euler equations
\begin{eqnarray}
M\frac{d{\bf V}}{dt} &=&\pmb{\mathcal{{F}}},\label{Eq:TranslationalVelocityOfSphere}\\
I\frac{d\mbox{\boldmath$\omega$}}{dt} &=&\pmb{\mathcal{T}},\label{Eq:RotationalVelocityOfSphere}
\end{eqnarray}
where $M$ and $I$ are the mass and moment of inertia of the spherical particle, and $\bf V$ 
and $\mbox{\boldmath$\omega$}$ are the translational and rotational 
velocities of the spherical particle, respectively. Equations (\ref{Eq:TranslationalVelocityOfSphere}) 
and (\ref{Eq:RotationalVelocityOfSphere}) can be solved for ${\bf V}$ and 
$\mbox{\boldmath$\omega$}$. The total velocity $\bf U$ that combines both the 
translational and rotational motion of the rigid body is given by 
${\bf U}={\bf V}+({\bf y} -{\bf X})\times \mbox{\boldmath$\omega$},\quad {\bf y}\in S$. 
Correspondingly, the equation of motion for points on the surface is 
\begin{eqnarray}
\frac{d{\bf y}}{dt} &=& {\bf U}\label{Eq:SphericalBodyPosition},\quad {\bf y}\in S. 
\end{eqnarray}

\subsection{Numerical approximation of the force and the torque within the DSMC scheme}

 To compute the force $\pmb{\mathcal{{F}}} $ and the torque $\pmb{\mathcal{T}}$ on the spherical particle, 
 at first we have to compute the stress tensor $\sigma$ in the fluid domain and finally insert the value 
 of $\sigma$ in the equations (\ref{Eq:HydrodynamicForceOnSphere}) and (\ref{Eq:HydrodynamicTorqueOnSphere}). 
For a dilute gas, the flow is modeled by  
 kinetic theory and given by the Boltzmann equation (\ref{Eq:Boltzmann}). The stress tensor $\sigma$ 
can be computed as a moment of the phase-space distribution function. However, for moving boundary 
problems  the correct numerical approximation of the stress tensor in the DSMC cells which are partially 
covered by the rigid body is not very accurate, because of the small  number of simulated molecules 
in that cell. Thus, we  compute the force and the torque from the interaction of the fluid molecules and the 
rigid particle. This is similar to a  microscopic approach where the force and the torque would be  
computed by the collision of the spherical particle with the simulated gas molecules. When the gas molecules 
collide with the spherical particle, they transfer momentum and energy.  
Therefore, the total force and the total torque exerted on the spherical particle are computed by accumulating 
the increments of the linear and angular momentum imparted by all the colliding 
molecules to the rigid body. This leads to  the following procedure to approximate numerically the force and the torque exerted  on the spherical particle.\\

Let us discretize the boundary $\partial S $ of the spherical particle by a uniformly distributed 
pointset $\partial S_{h}=\{{\bf y}_i, i=1,\ldots,  N_1\}$. Let us consider a gas molecule hitting the 
particle surface at a point ${\bf y}\in \partial S$ with momentum ${\bf p}$, being reflected with 
momentum ${\bf p}'$. In the following we assume that gas molecule is reflected diffusively from the boundary of the moving 
spherical particle. Then we find the closest neighbor ${\bf y}_i$ of ${\bf y}$ in the pointset $\partial S_{h}$ to store pre- and post- collision momenta of that molecule. During the 
time interval $\Delta t$, there could be a number of such gas molecules impinging at a position whose closest neighbor is ${\bf y}_i$. The total pre- and post- collision momenta 
at ${\bf y}_i$ are calculated by taking the sum of  pre-collision momenta ${\bf p}$ and  
post-collision momenta ${\bf p}'$ of those molecules which impinge in the 
neighborhood of ${\bf y}_i$. Let ${\bf p}_i$ and ${\bf p}'_i$ be the total pre- and post-collision 
momenta at the point ${\bf y}_i$. Then, the force ${\pmb{\mathcal{{F}}}}_i$ and the 
torque $\pmb{\mathcal{T}}_i$ exerted on the spherical particle at ${\bf y}_i$ during the time interval $\Delta t$ are given by
\begin{eqnarray}
{\pmb{\mathcal{{F}}}}_i &=& \frac{{\bf p}_i-{\bf p}'_i}{\Delta t},\quad i=1, \ldots,  N_1 \\
\pmb{\mathcal{T}}_i &=& ({\bf y}_i-{\bf X})\times{\pmb{\mathcal{{F}}}}_i, \quad i=1, \ldots,  N_1, 
\end{eqnarray}
where $\bf X$ is the center of mass of the spherical particle.
Hence, the total force $\pmb{\mathcal{{F}}}$ and the total torque $\pmb{\mathcal{T}}$ on the spherical particle are given by
\begin{eqnarray}
\pmb{\mathcal{{F}}} &=& \sum_i{\pmb{\mathcal{{F}}}}_i = \sum_i{\frac{{\bf p}_i-{\bf p}'_i}{\Delta t}} \label{Eq:ForceOnSphere1}\\
\pmb{\mathcal{T}} &=& \sum_i \pmb{\mathcal{T}}_i = 
\sum_i{({\bf y}_i-{\bf X})\times{\pmb{\mathcal{{F}}}}_i }\label{Eq:TorqueOnSphere1}. 
\end{eqnarray}
Each DSMC  molecule represents a  large number of physical gas molecules. Therefore, 
we need to determine the mass of each  DSMC simulated molecule to be used in the momentum transfer calculation. We employ the ideal gas law 
\begin{eqnarray}\label{Eq:IdeaGasEqn1}
pV &=& n \Re T \quad\textrm{rewritten as}\\\label{Eq:IdeaGasEqn2}
p &=& \rho R T, 
\end{eqnarray}
where $n$ is the number of moles and $V$ is the system volume. $\Re$ and $R$ 
are the universal and specific gas constants, respectively. 
From (\ref{Eq:IdeaGasEqn1}) and (\ref{Eq:IdeaGasEqn2}), we can write
\begin{eqnarray}
n &=& \frac{\rho V}{m_\textrm{g} N_\textrm{\tiny A}}, 
\end{eqnarray} 
where $N_\textrm{\tiny A}$ is the Avogadro  number. The total number of 
physical gas molecules in the system is 
\begin{equation}\label{Eq:NumnerOfMolecules}
N=\frac{\rho V}{m_\textrm{\tiny g}},
\end{equation} 
where $m_\textrm{g}$ is the mass of a physical gas molecule.\\
 
Define
\begin{equation}\label{Eq:WeightSimulatedMolecule}
\nu := \frac{N}{N_0}. 
\end{equation} 
Here, $N_0$ is the total number of DSMC simulated molecules, 
hence $\nu\ge 1$ is the number of physical gas molecules representing a single DSMC 
simulated molecule. This is also known as the statistical weight of the simulated DSMC molecule. 
Thus, the mass of a DSMC molecule is given by
\begin{equation}\label{Eq:MassSimulatedMolecule}
M = m_\textrm{g}\nu. 
\end{equation}
Equations (\ref{Eq:NumnerOfMolecules}), (\ref{Eq:WeightSimulatedMolecule}) 
and (\ref{Eq:MassSimulatedMolecule}) finally yield
\begin{equation}
M = \frac{\rho V}{N_0}.
\end{equation}
Thus, the force and the torque  (\ref {Eq:ForceOnSphere1}) and (\ref {Eq:TorqueOnSphere1}) are rewritten as
\begin{eqnarray}
\pmb{\mathcal{{F}}} &=& \sum_i{\frac{M({\bf v}_i-{\bf v}'_i)}{\Delta t}}, \label{Eq:ForceOnSphere2}\\
\pmb{\mathcal{T}} &=& \sum_i{({\bf y}_i-{\bf X})\times{\frac{M({\bf v}_i-{\bf v}'_i)}{\Delta t}}},
\label{Eq:TorqueOnSphere2}
\end{eqnarray}
where ${\bf v}_i$ and ${\bf v}'_i$ are the pre- and post- collisional velocities. 
With these expressions we can determine the translational and rotational motion 
of the spherical particle, solving    
(\ref{Eq:TranslationalVelocityOfSphere}), (\ref{Eq:RotationalVelocityOfSphere}) and (\ref{Eq:SphericalBodyPosition}). 
For the time integration we use an explicit Euler scheme with given initial velocities 
${\bf V}(t=0) ={\bf V}_0,~ \mbox{\boldmath$\omega$}(t=0)=\mbox{\boldmath$\omega$}_0$ 
and initial configuration $S(0) = S_0$ of the spherical particle.   We have considered the 
same time step $\Delta t$ for both Newton-Euler equations as well as the DSMC method.  
 
\section{Brownian motion}
The theory of translational  Brownian motion is concerned with the calculation of the 
probability density for the position of a rigid particle in a fluid. It  is usually based on  
Langevin's equation, which is Newton's second law with the assumption that the 
force acting on the rigid particle is the sum of a viscous retarding force proportional to 
the velocity of the rigid particle and a rapidly fluctuating force whose statistical properties 
are such that the the velocity distribution approaches a Maxwell-Boltzmann 
distribution \cite{hubbard,nelson}. 
Analogously, rotational Brownian motion is concerned with the calculation of the 
probability density of the orientation of a body in a fluid.  The specification 
of the orientation of a body requires three coordinates, such as Euler's angles. 
Here, we  consider the simple case where the rigid particle rotates about  a 
fixed axis through its center of mass. 
The theory of rotational Brownian motion is again based on a Langevin equation. 

 \subsection{Translational Brownian motion of a rigid particle }
Consider a system composed of $N$  monoatomic  gas molecules  
occupying a volume $V$ and having an absolute temperature  $T$. 
Let us consider a rigid particle suspended in a rarefied gas. The particle undergoes a random motion due to the impacting gas molecules.  
Let  ${\bf V} = \frac{d{\bf X}(t)}{dt}$ denote  the velocity of the particle relative to the gas. 
The theory of Brownian motion states that the velocity of the particle follows the Maxwell-Boltzmann velocity distribution which is given by \cite{wax}
\begin{equation}\label{Eq:MaxwellVelDisParticle}
f_{ \bf V} = (\frac{M}{2\pi k_\textrm{\tiny B} T})^{3/2} \mbox{exp} \Big(-\frac{M\vectornorm{\bf V}^2}{2k_\textrm{\tiny B}T}\Big), 
\end{equation}
where $ M $ is the mass of the particle and  $k_\textrm{\tiny B}$ is the 
Boltzmann constant.  The mean square velocity can be calculated by taking  the 
second moment of the velocity distribution (\ref{Eq:MaxwellVelDisParticle}) and is given by 
\begin{equation}\label{Eq:MeanSquareVelParticle}
<\vectornorm{\bf V}^2>=\int_{\mathbb{R}^3} \vectornorm{\bf V}^2 f_{\bf V} d{\bf V}=\frac{3k_\textrm{\tiny B}T}{M}. 
\end{equation} 
Hence, the mean translational kinetic energy of the particle is given by
\begin{equation}\label{Eq:TransKeParticle}
\frac{1}{2}M<\vectornorm{\bf V}^2>=\frac{3}{2}k_\textrm{\tiny B}T. 
\end{equation}
The random impact of the surrounding gas molecules  generally causes two kinds of 
effects: firstly, they act as a random driving force on the Brownian particle to maintain 
its  irregular motion, and, secondly, they give rise to a friction force.  

A simple model for the Brownian motion of a particle with mass $M$ 
and center of mass ${\bf X}$ is the phenomenological 
stochastic equation \cite{kubo}
\begin{equation}\label{Eq:Langevin1}
M\frac{d^2{\bf X}(t)}{dt^2}=-\gamma \frac{d{\bf X}(t)}{dt}+\varsigma {\bf F}(t),
\end{equation}
denoted as Langevin equation. The frictional force exerted by the medium is represented by the first term on 
 the right-hand side, where $\gamma$ is the translational friction coefficient,
 assumed to be independent of the particle velocity. The second term, $\varsigma {\bf F}(t)$, is the 
 random force due to collisions with the surrounding gas molecules. \\
For the sake of simplicity and idealization, the random force is usually 
assumed to be a white noise process \cite{kubo} with autocorrelation function 
\begin{equation}\label{Eq:AutocorrelationForce}
 <{\bf F}(t_1){\bf F}(t_2)>=\delta(t_1-t_2)\mathbb{I}.
\end{equation}

At long times, the explicit solution of equation (\ref{Eq:Langevin1}) yields a variance  for the velocity 
${\bf V} = \frac{d{\bf X}(t)}{dt}$ given by
$$\frac{3 \varsigma}{2 \gamma M}.$$

Assuming that the distribution of velocites follows a Maxwell-Boltzmann distribution, we obtain (see Nelson \cite{nelson})
\begin{equation}\label{Eq:FluctuationDissipationTheorem}
\varsigma = 2 \gamma k_\textrm{\tiny B} T.
\end{equation}
For long times, the corresponding variance of ${\bf X}(t)$ is 
given by $6 D_T t $, with the translational diffusion coefficient
\begin{equation}
\label{Eq:DiffusionCoeff2}
D_T= \frac{k_\textrm{\tiny B} T }{\gamma},
\end{equation}
resulting in
\begin{equation}\label{Eq:Langevin7}
<\vectornorm{{\bf X}(t)-{\bf X}_0}^2>=\frac{6k_\textrm{\tiny B}T}{\gamma}t, 
\end{equation}
where ${\bf X}_0={\bf X}(t=0)$ is the initial position of the particle.
This is the same variance  as for a classical diffusion process with 
diffusion constant $D_T$.
(\ref{Eq:Langevin7}) is usually referred to as the Einstein equation. We note that relation (\ref{Eq:FluctuationDissipationTheorem})
 is an explicit manifestation of the fluctuation-dissipation theorem \cite{kubo}. It sets a constraint to the random force whose power spectrum is determined by the level of friction.

In Stokes flow regime, assuming small Knudsen 
and Reynolds numbers, the friction coefficient of a spherical 
particle has the following form \cite{stokes}
\begin{equation}\label{Eq:StokesFrictionCoeffSphere}
\gamma_\textrm{\tiny C}=6\pi\mu R_\textrm{\tiny P}, 
\end{equation} 
where $\mu$ is the viscosity of the fluid and $R_\textrm{\tiny P}$ is the radius of 
the spherical particle. This equation is valid under the assumption that the 
fluid satisfies a no-slip boundary condition, meaning that the relative velocity of 
the fluid at the solid surface is zero. This assumption holds in the continuum regime where the Knudsen number $Kn<<1$. Substituting the 
value of the friction coefficient $\gamma_\textrm{\tiny C}$ 
from (\ref{Eq:StokesFrictionCoeffSphere}) to  (\ref{Eq:DiffusionCoeff2}), we get 
\begin{equation}\label{Eq:StokesEinsteinRelation}
D_{\textrm{\tiny T,C}}=\frac{k_\textrm{\tiny B} T}{6\pi \mu R_\textrm{\tiny P}}. 
\end{equation}
This is known as the Stokes-Einstein equation for diffusion of spherical 
particles in  a fluid.

 In the case of a very large Knudsen number $Kn>>1$, an expression for the friction coefficient was derived by 
 Epstein\cite{epstein} using kinetic theory
 \begin{equation}\label{Eq:FreeMolTranslationalFrictionCoeff}
 \gamma_\textrm{\tiny FM}=\frac{8}{3}R_\textrm{\tiny P}^2\rho \sqrt{\frac{2\pi k_\textrm{\tiny B} T}{m_\textrm{g}}}\Big(1+\frac{\pi\alpha}{8}\Big), 
 \end{equation}
 where $\rho$ is the density of the gas. The coefficient $\alpha$ represents the fraction of gas molecules 
 that are reflected diffusively, $(1-\alpha)$ is the fraction of molecules with 
 specular reflection. \\
The translational diffusion coefficient of a spherical particle in a rarefied gas is obtained 
by combining  equations (\ref{Eq:DiffusionCoeff2}) 
and (\ref{Eq:FreeMolTranslationalFrictionCoeff}) and is given by
 \begin{equation}\label{Eq:DiffusionCoeff3}
 D_{\textrm{\tiny T,FM}}=
 \frac{3}{8}\sqrt{\frac{m_\textrm{g}k_\textrm{\tiny B}T}{2\pi}}\frac{1}{(1+\frac{\alpha \pi}{8})R_\textrm{\tiny P}{^2}\rho}. 
 \end{equation}
 
\subsection{Rotational Brownian motion and Brownian diffusion}
There  also exists rotational diffusion, the change of the orientation of a 
colloidal particle due to the random torque exerted on it by the surrounding molecules. Rotational diffusion is important, for example,  for the study of 
dielectric relaxation, fluorescence depolarization, or the line width in nuclear magnetic resonance measurements \cite{mazo}. 
The dynamics of rotational diffusion of aspherical  particle rotating about a fixed 
axis can be derived from the rotational analog of Langevin's equation based on the 
Euler equation \cite{hubbard}. This is given by
\begin{equation}\label{Eq:EulerRotationalMotion}
I \frac{d^2{\Theta}}{dt} = -\Upsilon{ \omega_3} + \Gamma,
\end{equation}  
where $\Theta$ is the angular displacement about a fixed axis parallel to the $z$-axis through 
the center of mass of the particle, ${\omega_3=\frac{d\Theta}{dt}}$ is the third 
component of the angular velocity $\boldsymbol{\omega}=(\omega_1,\omega_2,\omega_3)$, 
and  $I=\frac{2}{5}MR^2$ is the moment of inertia of the spherical particle. When the body 
rotates only about a fixed axis, the first two components $\omega_1,\omega_2$ of 
angular velocity $\boldsymbol{\omega}$ are zero.
The first term on the right hand side of (\ref{Eq:EulerRotationalMotion}) is the friction 
torque with rotational friction coefficient $\Upsilon$. The second term is the stochastic 
torque. This random torque has a correlation function given by  \cite{hubbard}  
\begin{equation}
 <\Gamma(t_1) \Gamma(t_2)> = 2\pi k_\textrm{\tiny B} T\Upsilon\delta(t_1-t_2).
\end{equation}
The amplitude $2\pi k_\textrm{\tiny B} T\Upsilon$ of the stochastic torque is given by 
similar considerations as before: By virtue of equipartition of energy, the distribution 
of the angular velocity of the spherical particle has to be a Maxwell-Boltzmann 
distribution given by \cite{hubbard}, \cite{wax}
\begin{equation}\label{Eq:MaxwellAngularVelDistributionParticle}
f_{\mbox{\boldmath$\omega$}} = \Big(\frac{I}{2\pi k_\textrm{\tiny B} T}\Big)^{3/2}
\mbox{exp} \Big(\frac{-I \vectornorm{\mbox{\boldmath$\omega$}}^2}{2k_\textrm{\tiny B}T}\Big).
\end{equation}
Therefore, the mean square angular velocity of the particle is given by
\begin{equation}\label{Eq:MeanSquareAngularVelocity}
<\vectornorm{\mbox{\boldmath$\omega$}}^2>=\int_{\mathbb{R}} \vectornorm{\mbox{\boldmath$\omega$}}^2f_
{\mbox{\boldmath$\omega$}}d{\mbox{\boldmath$\omega$}}=\frac{k_\textrm{ \tiny B}T}{I}.
\end{equation}
Hence, the rotational kinetic energy of the particle is 
\begin{equation}\label{Eq:RotationalKE}
\frac{1}{2}I < \vectornorm{\mbox{\boldmath$\omega$}}^2> =\frac{3}{2}k_\textrm{\tiny B}T.
\end{equation}
Since  the components $\omega_1,\omega_2,\omega_3$ of the rotational velocity of the 
Brownian particle are mutually independent, the rotational kinetic energy in each component is 
\begin{equation}\label{Eq:RotationalKEachComponent}
\frac{1}{2}I <\omega_i^2> =\frac{1}{2}k_\textrm{\tiny B}T,~i=1,2,3. 
\end{equation}

The theory of rotational diffusion can be derived in analogy to the theory of translational diffusion. 
For large times the solution of equation (\ref{Eq:EulerRotationalMotion}) can be approximated 
by the solution of a diffusion equation with rotational diffusion coefficient
\begin{equation}\label{Eq:RotationalDiffusion1}
D_\textrm{\tiny R}=\frac{k_\textrm{\tiny B} T}{\Upsilon}. 
\end{equation}

Thus, if at time $t=0$ the orientation of the particle is $\Theta = \Theta_0$, the   mean 
square angular displacement of the particle is given by
\begin{eqnarray}
<\lvert \Theta-\Theta_0\rvert^2> &=& 2 D_\textrm{\tiny R} t \label{Eq:MeanSquareAngularDisp}. 
\end{eqnarray}

 The rotational friction coefficient is the proportionality constant between the drag torque on the 
 particle and its angular velocity about a fixed axis. 
 In the continuum regime $(Kn<<1)$, the rotational friction coefficient of a spherical 
 particle is given by \cite{loyalka}
\begin{equation}\label{Eq:RotationalFricCont}
\Upsilon_\textrm{\tiny C} = 8\pi\mu R_\textrm{\tiny P}^3. 
\end{equation}
It depends on the viscosity $\mu$ of the fluid and the radius of the particle $R_\textrm{\tiny P}$.\\
In free molecular regime $(Kn>>1)$, the expression for the rotational friction 
coefficient is \cite{loyalka,epstein}
\begin{equation}\label{Eq:RotationalFricFreeMol}
\Upsilon_\textrm{\tiny FM} =
 \frac{2\pi}{3}\sqrt{\frac{8k_\textrm{\tiny B} T}{\pi m_\textrm{g}}}\rho R_\textrm{\tiny P}^4. 
\end{equation} 
It depends, among others, on the temperature $T$, and density $\rho$ of the gas. 
Correspondingly, in the free molecular regime,  the rotational diffusion coefficient is given by
\begin{equation}\label{Eq:RotationalDiffusion2}
D_{\textrm{\tiny R,FM}} = \frac{3}{4\pi}\sqrt{\frac{\pi m_\textrm{g} k_\textrm{\tiny B} T}{2}}\rho R_\textrm{\tiny P}^4. 
\end{equation}
 
\section{Numerical results}

\begin{figure} 
\centering
\includegraphics[keepaspectratio=true, width=.48\textwidth]{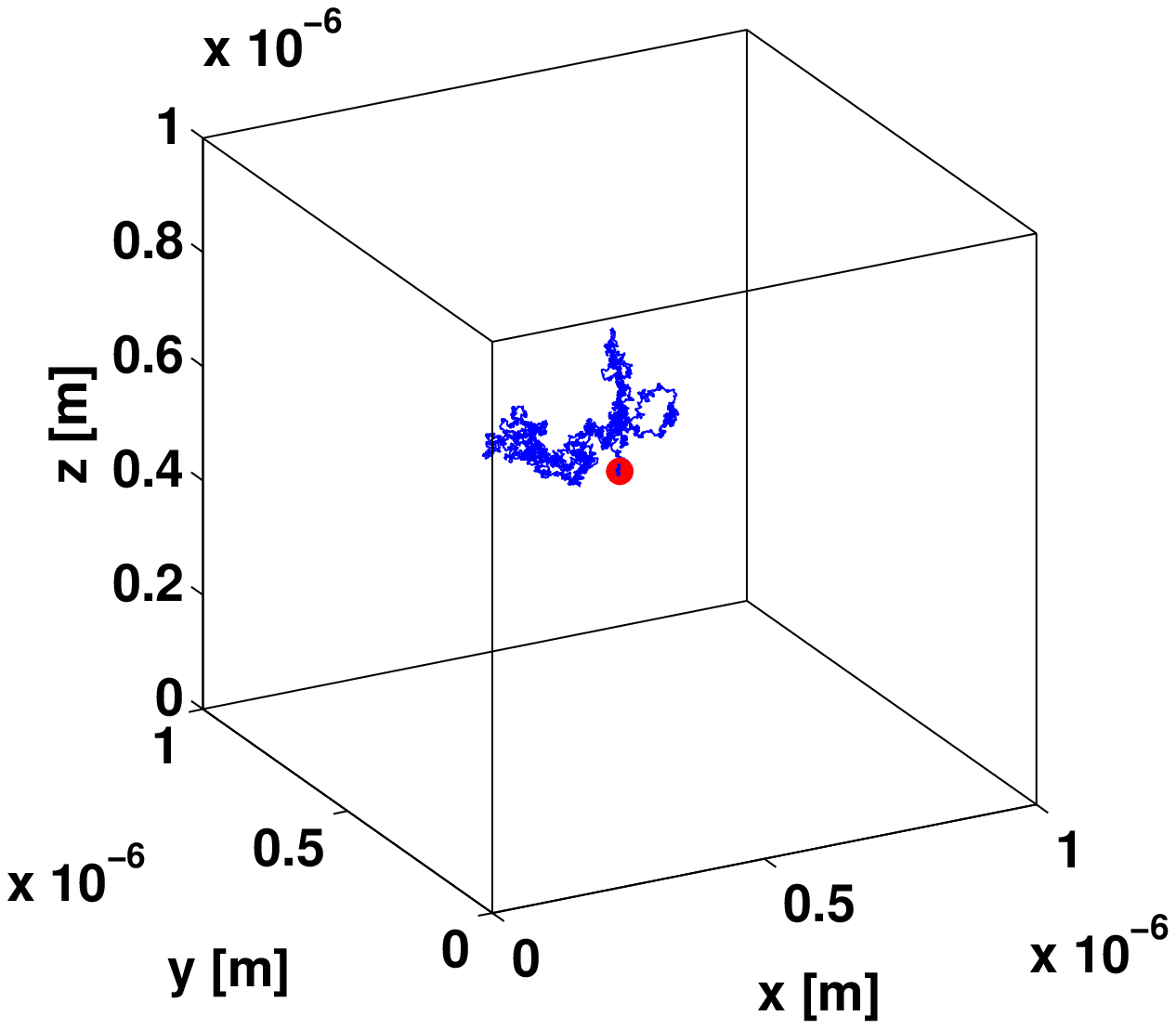}
 \includegraphics[keepaspectratio=true, width=.48\textwidth]{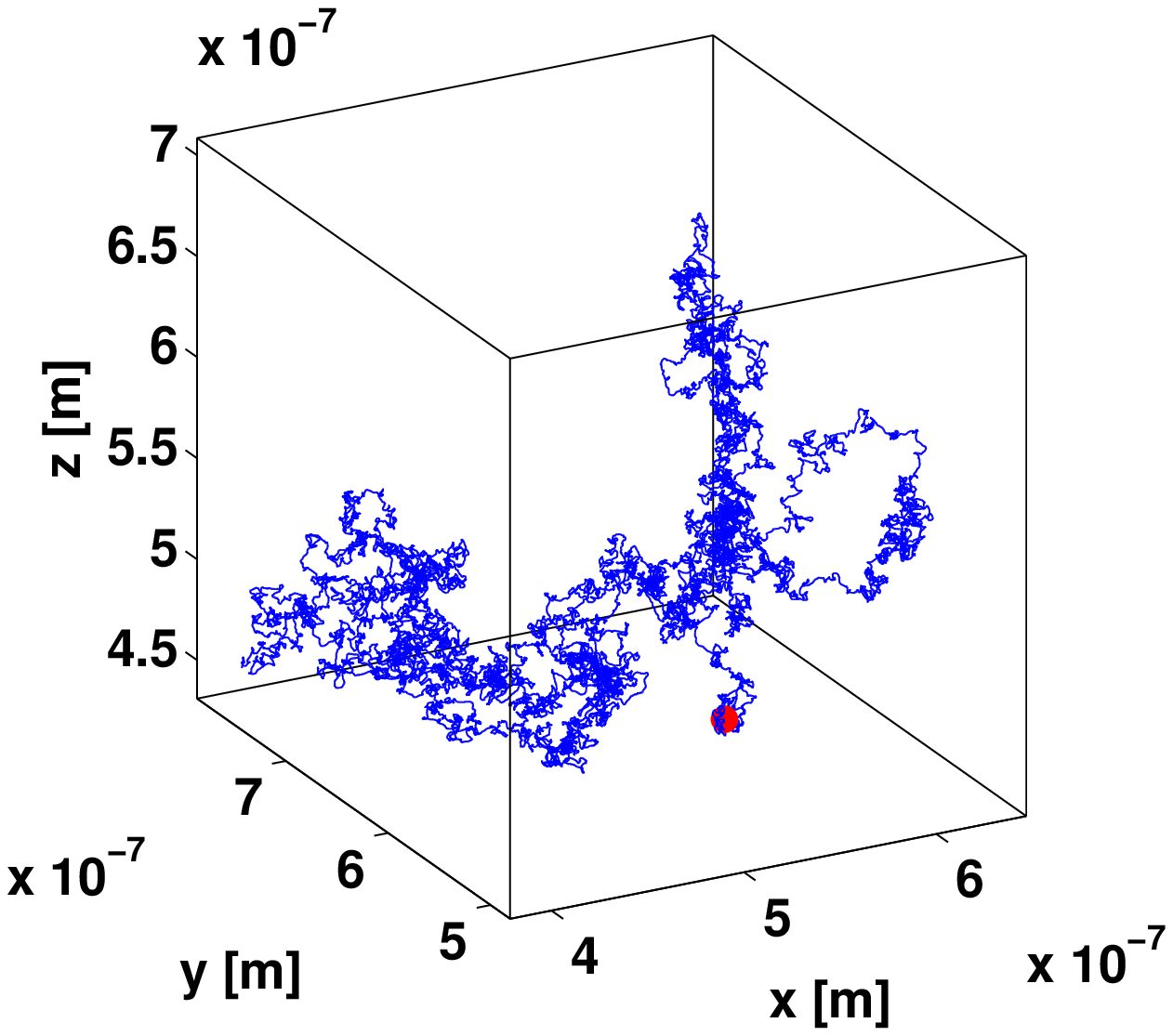}
\caption{Brownian motion of the spherical particle until a time of $3.7879\times 10^{-6}$ seconds.Left: normal view, right:  zoom view. }
\label{Fig:3DBrownainTrajectory}
\end{figure}

In this section, we present results for the above described numerical algorithm. 
The numerical method is validated by comparing the numerical approximation of
 the translational and rotational diffusion coefficients with the explicit formulas given in the previous section.
The cube-shaped computational domain is of size $ 10^{-6}\times 10^{-6}\times 10^{-6} ~ \mbox{m}^3$, the gas 
is kept at a uniform temperature of $T=300$~K. A uniform Cartesian grid of $30\times30\times30$ cells is defined inside the computational domain. 
We have considered different numbers $n_0$ of 
gas molecules (such that $\nu\ge 1$) to compute the numerical value of the diffusion 
coefficients and compare the results with the theoretical values. The radius of the 
spherical particle is taken to be $10^{-7}$~m.  The gas is argon, a 
monoatomic gas with mass $m_\textrm{g} = 6.63\times 10^{-26}$~kg,  
the Boltzmann constant is  $k_\textrm{\tiny B}=1.38\times 10^{-23}$~J/K, and  the 
specific gas constant is $R={k_\textrm{\tiny B}}/{m_\textrm{g}}=208$~J/(kgK) \cite{TKHD}. 
We use a  hard-sphere collision model with  diameter $d=3.68\times10^{-10}$~m. 
The numerical computation is performed  for a Knudsen number $Kn=11$, using the 
the particle diameter as length scale. The spherical particle is initially kept at rest  in the center of the 
computational domain, with its center of mass at $(5\times10^{-7} \text{m},5\times10^{-7} \text{m}, 5\times10^{-7} \text{m})$. 
Hard collisions are  performed between  gas molecules and the spherical particle. In our simulations, diffuse 
boundary conditions are applied at the boundary of the particle as well as at the walls of the domain. 
The force and the torque are computed using (\ref{Eq:ForceOnSphere2}) and (\ref{Eq:TorqueOnSphere2}). 
The trajectory of the spherical particle is computed using (\ref{Eq:SphericalBodyPosition}).   
The trajectory of the center of mass of the particle up to a time of $3.7879\times 10^{-6}$ seconds is 
shown in Fig. \ref{Fig:3DBrownainTrajectory}.\\

\begin{figure} 
\centering
\includegraphics[keepaspectratio=true, width=.48\textwidth]{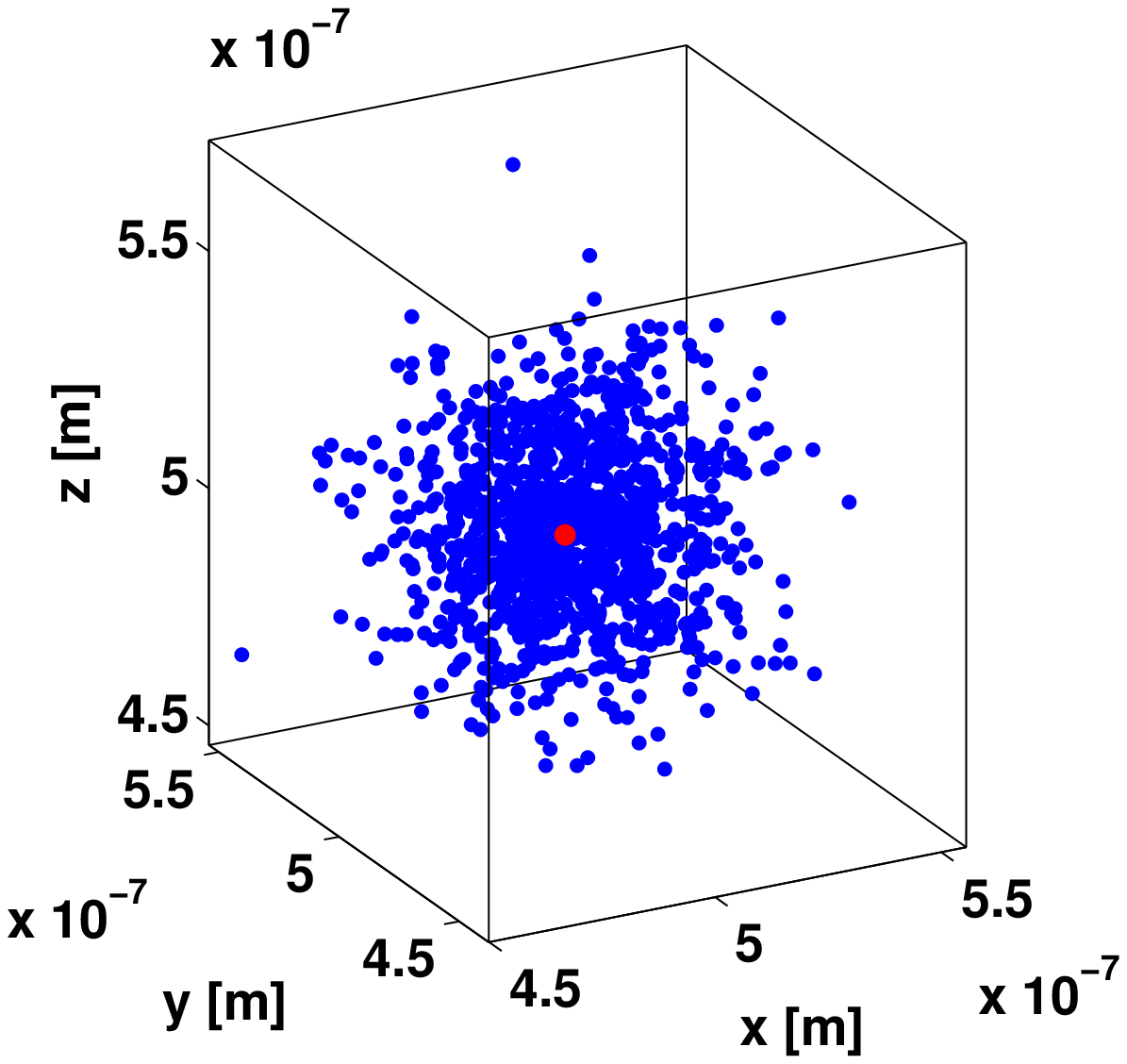}
\includegraphics[keepaspectratio=true, width=.48\textwidth]{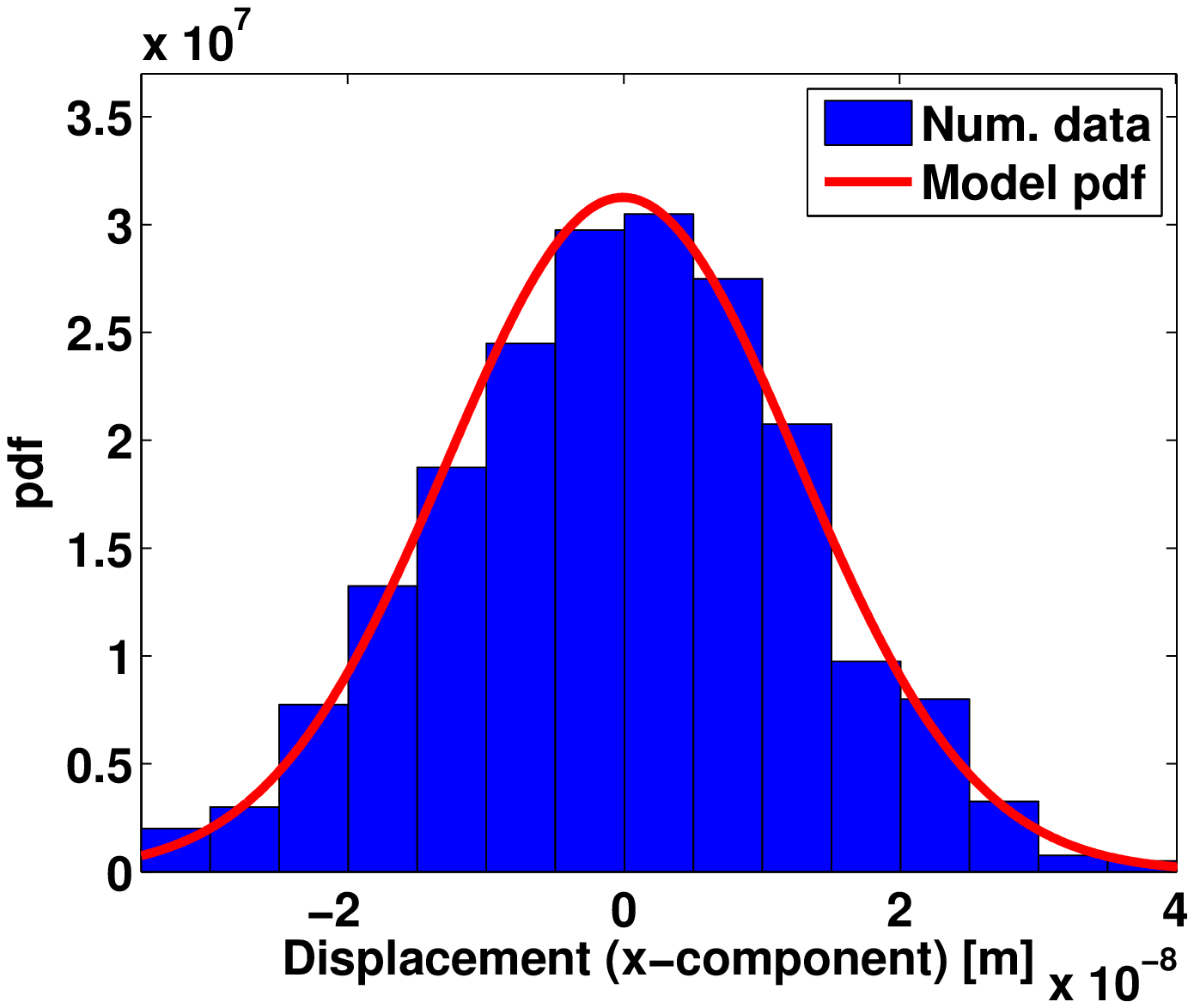}
\caption{ Endpoints of the Brownian trajectories of the spherical particles (left).  The histograms represent the simulation data and the solid line is the Gaussian distribution 
of the $x$-displacements (right), both at time $2.2617\times 10^{-8}$ seconds.}
\label{Fig:3DBrownainDiffusion}
\end{figure}

To compute the numerical translational diffusion coefficient of the particle, we have performed 
the experiment a number of times under similar physical conditions. The displacement of the center of 
particle is sampled at $t=2.617\times10^{-8}$ seconds. The corresponding distribution of endpoints of the particle 
trajectories is shown in figure  \ref{Fig:3DBrownainDiffusion} (left). 
The corresponding distribution of the  $x$-component of the displacement of the center of mass  is displayed in 
Fig.  \ref{Fig:3DBrownainDiffusion}  (right). The other components show a 
similar behaviour.  The center of mass of the particle follows a Gaussian 
distribution \cite{uhlenbeck} with zero mean displacement and a variance in each 
direction  approximately equal to $2D_\textrm{\tiny T}t$.

\begin{figure}
\centering
\includegraphics[keepaspectratio=true, width=.6\textwidth]{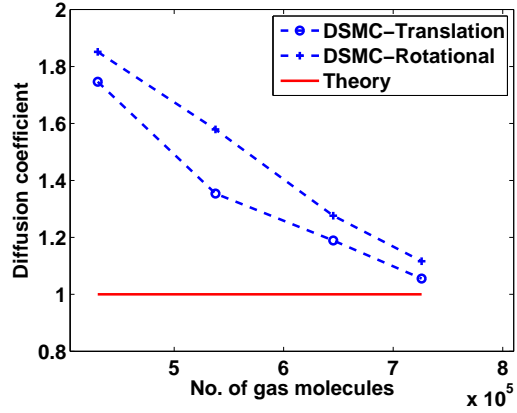}
\caption{Convergence  of the normalized  translational and rotational diffusion coefficients with increasing number of gas molecules. The solid line is the theoretical value, lines with - - o - - and - - + - -  represent the translational and the rotational diffusion coefficients, respectively.}
\label{Fig-3D_ConvergencePlotTransBrownianDiffusion}
\end{figure}

We have performed  numerical experiments for an increasing  number of simulated 
gas molecules and sampled the data to obtain the center of 
mass of the spherical particles at time $t=2.617\times10^{-8}$ seconds and the 
translational diffusion coefficient, computed by using (\ref{Eq:Langevin7}). 
The theoretical value of the translational diffusion coefficient is given by (\ref{Eq:DiffusionCoeff3}). Fig. \ref{Fig-3D_ConvergencePlotTransBrownianDiffusion} 
shows that the numerical approximation of the translational diffusion coefficient of the particle 
converges to the theoretical value for an increasing number of simulated molecules.

 The rotational diffusion coefficient of the spherical particle  in  equation (\ref{Eq:RotationalDiffusion2}) is derived for a non-moving particle which is  rotating only about a fixed axis through its center of mass. Thus, in our numerical simulation, we compute the value of the rotational diffusion coefficient of the spherical particle by fixing its 
center of mass and  letting the spherical particle rotate only around a single axis parallel to the $z-$ axis  by putting first and second components of the angular velocity equal to zero. 
 The torque is 
computed by using (\ref{Eq:TorqueOnSphere2}). As a result, one obtains the 
angular velocity $\mbox{\boldmath$\omega$}=(0,0,\omega_3)$.  
Finally, the angular displacement $\Theta$ is computed using the  equation of angular motion 
\begin{equation}\label{Eq:AngularDisplacement}
\frac{d\Theta}{dt}=\omega_3, \quad \textrm{with initial angle} \quad \Theta_0=0. 
\end{equation}

\begin{figure} 
\centering
\includegraphics[keepaspectratio=true, width=.6\textwidth]{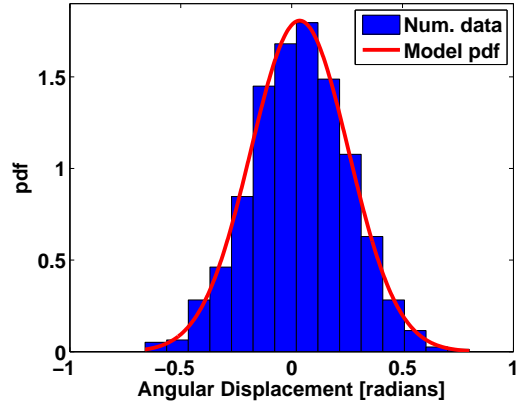}
\caption{ The histograms represents the simulation data and the solid line is the Gaussian distribution function of the angular displacement about the  $z$-axis at time $2.2617\times 10^{-8}$ seconds.}
\label{Fig:Pdf3DAngularBrownainDiffusion}
\end{figure}

The angular  displacement of the spherical particle is sampled at  
time $t=2.2617\times 10^{-8}$ seconds. The corresponding distribution function is shown in Fig.  {\ref{Fig:Pdf3DAngularBrownainDiffusion}}. 
It is well described by a Gaussian distribution with zero mean angular displacement and a variance of $2D_\textrm{\tiny R}t$.

We have also computed the rotational diffusion coefficient of the spherical particle for different numbers of simulated molecules by using (\ref{Eq:MeanSquareAngularDisp}). The approximated values are compared with the theoretical value of the rotational diffusion coefficient (\ref{Eq:RotationalDiffusion2}). The comparison of the numerically 
computed  values of the rotational diffusion coefficient with increasing number of simulated 
gas molecules and the theoretical value is again shown in Figure  \ref{Fig-3D_ConvergencePlotTransBrownianDiffusion}.

\begin{figure}  
\centering
 \includegraphics[keepaspectratio=true, width=.48\textwidth]{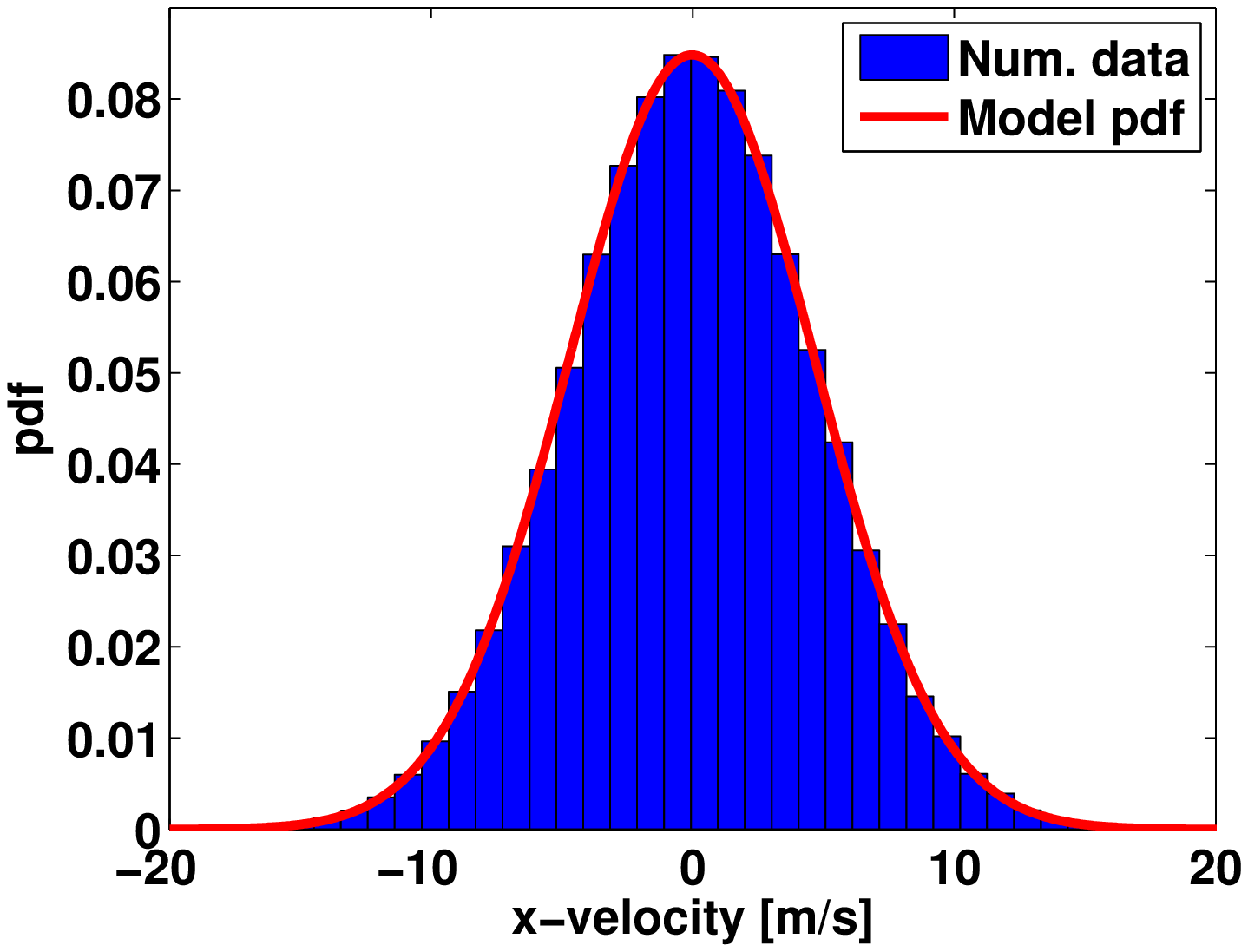}
 \includegraphics[keepaspectratio=true, width=.48\textwidth]{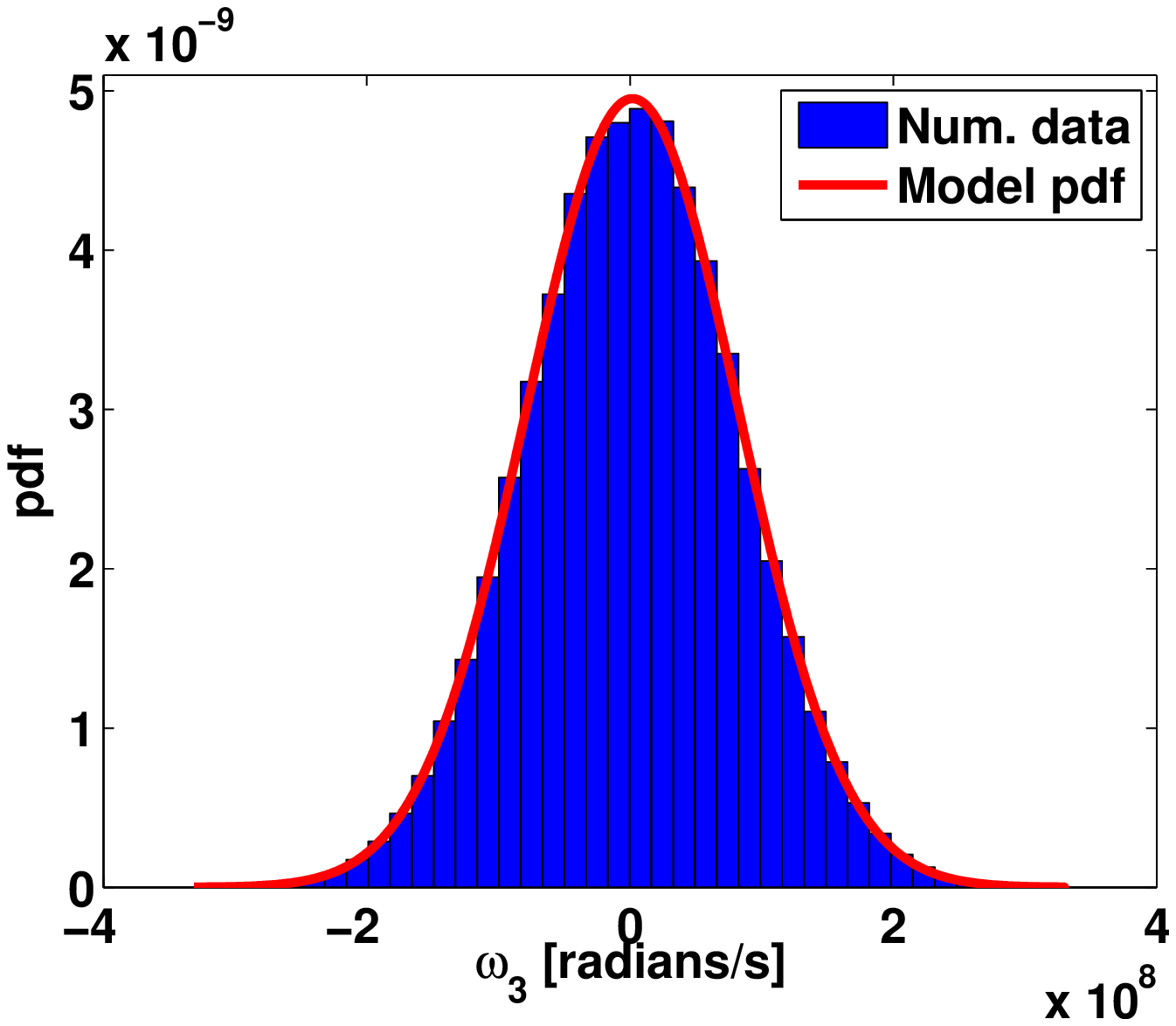}
 \caption{ x-component of the translational velocity (left), rotational velocity $\omega_3$ of the spherical particle (right). 
In both  figures the   histograms represent the simulation data and the solid line is the Gaussian distribution  function  at T = 300~K. }
\label{Fig:3DBrownainDiffusionTranslationalVelocityPdfPlot}
\end{figure}

For the Brownian motion of the colloidal particle  the translation and rotational velocities 
must follow the Maxwellian distributions  (\ref{Eq:MaxwellVelDisParticle}) 
and (\ref{Eq:MaxwellAngularVelDistributionParticle}). To test this, the numerical experiment has been run for a long time at a given temperature of $T=300~{\text K}$, and the translational and rotational velocities have been sampled at each time step. 
Figs.  (\ref{Fig:3DBrownainDiffusionTranslationalVelocityPdfPlot})  show the distribution of the translational velocity in $x$-direction and the distribution of the rotational velocity. The histograms represent 
the simulated data, and the solid lines are the model Gaussian curves. It can be concluded that the numerically computed probability density functions agree with the velocity distributions (\ref{Eq:MaxwellVelDisParticle}) 
and (\ref{Eq:MaxwellAngularVelDistributionParticle}).

\section{Conclusion and outlook}

We have presented a numerical method for simulating rigid bodies
with arbitrary shape  in a rarefied gas. 
The rarefied gas is modeled by the Boltzmann equation, and the motion of the particle is 
described by the Newton-Euler equations. The Boltzmann equation is solved by a DSMC 
type of particle method with a hard-sphere collision term. 
The forces acting on the particle are computed from the momentum transfer due to the molecules colliding with the surface.  
Translational and rotational diffusions of a spherical particle was investigated numerically. 
The obtained probability density functions for the particle's center-of-mass position, the translational and rotational velocities, and 
the corresponding diffusion coefficients were compared with results from the theory of Brownian motion. 
The convergence of the diffusion coefficients to their theoretical values with increasing number of 
simulated molecules was demonstrated. Future work will concentrate on computing 
thermophoretic transport processes of (non) spherical particles at different Knudsen numbers using the approach presented here.

\subsection*{\bf Acknowledgment}  
\noindent
This  work is partially supported 
by the German research foundation, DFG grant KL
1105/20-1 and by the DAAD PhD programme MIC.


\begin{thebibliography}{Duli76}
 
\bibitem{BI89} H. Babovsky, R. Illner, { A Convergence Proof for Nanbu's Simulation 
Method for the Boltzmann Equation}, {\it SIAM J. Numer. Anal.}, 26 (1989) 45-64. 
 
\bibitem{balakrishnan} K. Balakrishnan, J. B. Bell, A. Donev and A. L. Garcia 
{Fluctuating Hydrodynamics and Direct Simulation Monte Carlo}, 28th International 
Symposium on Rarefied Gas Dynamics, {\it AIP Conf. Proc.}, 1501(2012), 695-704.
\bibitem{batchelor} G. K. Batchelor, {An Introduction to Fluid Dynamics}, Cambridge University Press, 1967.
\bibitem{bird} G. A. Bird, {Molecular Gas Dynamics and Direct Simulation of Gas Flows}, Clarendon, Oxford, 1994.
 
\bibitem{cercignani1} C. Cercigani, { The Boltzmann Equation and its Applications}, Springer, Berlin, 1988.
\bibitem{cercignani2} C. Cercignani, M. Lampis, { Kinetic Model of Gas-surface Interactions}, {\it Transp. Th. and 
Stat. Phys.}, 1 (2) (1971) 101-114.

\bibitem{DM} G. Dechriste', L. Mieussens, Numerical simulation of micro flows with moving obstacles, {\it J. Phys.: Conf. Series} 362 (2012) 012030. 
 
\bibitem{epstein} P. S. Epstein, {On the Resistance Experienced by Spheres in Their Motion Through Gases}, {\it Phys. Rev.} 23 (1924) 710.
 
\bibitem{GRT}
M. A. Gallis, D. J. Rader, and J. R. Torczynski,
Thermophoresis in Rarefied Gas Flows, {\it 
Aerosol Science and Technology},  36 (2002) 1099-1117. 

 \bibitem{hubbard} P. S. Hubbard, {Rotational Brownian Motion}, {\it Phys.  Rev. A}, 6 (1972) 2421.
 
 \bibitem{KBA} G. Karniadakis, A. Beskok, N. Aluru, { Microflows and Nanoflows: Fundamentals 
  and Simulations}. Springer., New York, 2005.  
   
\bibitem{kubo} R. Kubo, { The Fluctuation-dissipation Theorem}, {\it Rep. Prog. Phys.}, 29 (1966) 255.
 
 \bibitem{loyalka} S. K. Loyalka, Motion of a Sphere in a Gas: Numerical Solution of the Linearized Boltzmann Equation, {\it Phys. Fluid A}, 4 (1992) 1049.

\bibitem{maedler} L. M$\mathrm{\ddot{a}}$dler, S.K. Friedlander, {Transport of Nanoparticles in 
Gases: Overview and Recent Advances}, {\it Aerosol and Air Quality Research}, 7(3) (2007) 304-342.
\bibitem{mazo} R. M. Mazo, {Brownian Motion: Fluctuations, Dynamics, and Applications}, Oxford University Press, 2002.

\bibitem{nelson} E. Nelson { Dynamical theories of Brownuan motion}, Princeton University Press, 2nd Ed. , 2001. 
 
\bibitem{NS95}H. Neunzert, J. Struckmeier, {Particle Methods for the Boltzmann Rquation}, {\it Acta Numerica}, 4 (1995), 417-457. 
 
\bibitem{philipse} A.P. Philipse, { Notes on Brownian Motion}, Utrecht University, Debye Institute, Van 't Hoff Laboratory, August 2011.

\bibitem{russel} W.B. Russel, {Small Particles Suspended in Liquid}, {\it Ann. rev. Fluid Mech.}, 13 (1981) 425-55.

\bibitem{RF} G. Russo, F. Filbet, Semi-Lagrangian schemes applied to moving boundary problems for 
the BGK model of rarefied gas dynamics, {\it Kinetic and Related Models} 2 (2009), 231-252. 
 
\bibitem{stokes} G. G. Stokes, {On the Effect of Fluids on the Motion of Pendulums},{ \it Trans. Cambridge Philos. Soc.}, 9, 8 (1851). 
Reprinted in Mathematical and Physical Papers III (Cambridge University Press, Cambridge, MA).

\bibitem{TA}T. Tsuji, K. Aoki, Moving boundary problems for a rarefied gas: Spatially one dimensional case, {
\it J. Comp. Phys.} 250 (2013), 574-600. 

\bibitem{TKH} S. Tiwari, A. Klar, S. Hardt, A particle-particle hybrid method for kinetic and continuum equations, 
{\it J. Comput. Phys.} 228 (2009) 7109-7124. 

\bibitem{TKHD} S. Tiwari, A. Klar, S. Hardt, A. Donkov, Simulation of a Moving Liquid Droplet Inside a Rarefied 
Gas Region, {\it Computers \& Fluids.} 71 (2013) 283-196.
 
\bibitem{uhlenbeck} G. E. Uhlenbeck, L. S. Ornstein, {On the Theory of Brownian 
Motion}, {\it Physical Rev.}, 36 (1930). 
 \bibitem{wax} N. Wax, {{Selected Papers on Noise and Stochastic Processes}}, 
 edited by N. Wax, Dover, New York , 1954.

\end{thebibliography}
\end{document}